\documentclass[iop]{emulateapj}
\usepackage[latin9]{inputenc}
\usepackage{xcolor}
\usepackage{units}
\usepackage{graphicx}
\PassOptionsToPackage{normalem}{ulem}
\usepackage{ulem}
\usepackage[unicode=true,
 bookmarks=true,bookmarksnumbered=true,bookmarksopen=true,bookmarksopenlevel=1,
 breaklinks=false,pdfborder={0 0 0},pdfborderstyle={},backref=false,colorlinks=false]
 {hyperref}
\hypersetup{pdftitle={Implications of time-varying potentials for Jeans analyses},
 pdfauthor={Tim Haines, Elena D'Onghia, Benoit Famaey, Chervin Laporte, and Lars Hernquist}}
\usepackage{breakurl}


\begin{document}
\title[Vertically perturbed Galactic disk]{Implications of a time-varying Galactic potential \\ for determinations of the dynamical surface density} \author{Tim Haines\altaffilmark{1} and Elena D'Onghia\altaffilmark{1,*}} \email{thaines@astro.wisc.edu} \affil{Department of Astronomy, University of Wisconsin-Madison, 475 N. Charter Street, Madison, WI 53076, USA.} \author{Benoit Famaey} \affil{Universit\'e de Strasbourg, CNRS, UMR 7550, Observatoire astronomique de Strasbourg, 11 rue de l'Universit\'e, 67000, Strasbourg, France} \author{Chervin Laporte\altaffilmark{2}} \affil{Department of Physics \& Astronomy, University of Victoria, 3800 Finnerty Road, Victoria BC, V8P 5C2 Canada}
\and \author{Lars Hernquist} \affil{Harvard Smithsonian Center for Astrophysics, 60 Garden Street, Cambridge, MA 02138, USA} \altaffiltext{*}{Center for Computational Astrophysics, Flatiron Institute, 162 Fifth Avenue, New York, NY 10010, USA.}
\altaffiltext{2}{CITA National Fellow}

\begin{abstract}
We analyze a high-resolution N-body simulation of a live stellar disk perturbed by the recent passage of a massive dwarf galaxy that induces a wobble, in-plane rings and phase spirals in the disk. The implications of the phase-space structures and  the estimate of the matter density through traditional Jeans modeling are investigated. The dwarf satellite excites rapid time-variations in the potential, leading to a significant bias of the local matter surface density determined through such a method. In particular, while the Jeans modelling gives reasonable estimates for the surface density in the most overdense regions of the disk, we show that it fails elsewhere. 
{\bf Our results show that the spiral-shape feature in the $z-v_z$ plane is visible preferentially in the under-dense regions and vanishes more quickly in the inner parts of the stellar disk or in high-density regions of the disk}.
While this prediction can be verified with \textit{Gaia} DR3 our finding is highly relevant for future attempts in determining the dynamical surface density of the outer Milky Way disk as a function of radius. The outer disk of the Milky Way is indeed heavily perturbed, and \textit{Gaia} DR2 data have clearly shown that such phase-space perturbations are even present locally. While our results show that traditional Jeans modelling should give reliable results in overdense regions of the disk, the important biases in underdense regions call for the development of non-equilibrium methods to estimate the dynamical matter density locally and in the outer disk.

\end{abstract}

\section{Introduction}
 Thanks to the second Gaia data release (Gaia DR2) \citep{GaiaDR2}, a map of the kinematics of disk stars in the Galaxy can now be provided not only for the solar vicinity but also over a larger spatial extent out to a few kpc from the Sun, allowing us to explore the phase-space of the stellar disk. This will become even more topical when the Gaia DR2 will be combined with future large spectroscopic surveys such as 4MOST \citep{4MOST} and WEAVE \citep{WEAVE,WEAVE_SURVEY}.
 In principle, this should allow us to map the dynamical surface density as a function of height at various radii in the disk, and compare this with the baryonic density to infer the distribution of dark matter around the disk \citep[e.g.][]{bovy2013adirect}. Locally the dynamical surface mass density estimates inferred from stars oscillating vertically above and below the plane \citep{kuijken1991thegalactic,garbari2012anew,bovy2012onthe,bienayme2014weighing,HagenHelmi2018,kuijken1989themass,kuijken1989themass2,siebert2003vertical,holmberg2004thelocal,bovy2012onthe,Widmark2019,Widmark2} have been used to probe the local dark matter density, the dark matter vertical distribution and its possible deviations from a spherical halo profile, as well as the presence of a dark disk, thereby providing potential constraints on the nature of dark matter, and on alternatives \citep{bienayme2009}. In particular, if the dark matter density is very low at the solar vicinity, around the midplane, the relative contributions from visible and dark matter can be disentangled by trying to constrain the gravitational potential out to larger heights \citep{bovy2012onthe,holmberg2004thelocal}.

However, the various values reported for the dark matter density estimates in the solar neighborhood are affected 
by uncertainties that are associated with the large errors in the estimate of the gas mass density and the local stellar densities \citep[see][]{silverwood2016anonparametric}. Furthermore, all these analyses assume that the disk potential is static and in dynamical equilibrium. In particular potential collective modes in the Galactic disk are not accounted for. For example \citet{banik2017galactoseismology} recently studied models of a Galactic disk out of dynamical equilibrium to show that when a single
tracer population is adopted to measure the vertical force at the solar vicinity, the uncertainty in the inferred value may be of the order of 20\%.

\textit{Gaia} DR2 data indicate that the Milky Way stellar disk in the solar neighborhood is far from being smooth but shows evidence of substructure in the kinematic space that manifests in the form of ridges and arches. While such structures had long been known for planar motions \citep[e.g.][]{Dehnen1998,Famaey2005,Monari2017}, the finding of a phase-space spiral structure reported in the $z-v_z$ plane \citep{Antoja2018} could potentially have important implications for determinations of the dynamical surface density. This feature appears to extend well beyond the solar neighborhood  cylinder in which it was originally discovered and traced out to $R\sim11 \,\rm{kpc}$ \citep{Laporte2018} using Gaia DR2 \citep[see also][for a GALAH $R_{\odot}\pm0.5\,\rm{kpc}$ exploration]{BlandHawthorn2019}.

Previous observational facts already indicated that the stellar disk is out of dynamical equilibrium, namely the vertical bulk motion of stars discovered in the solar neighborhood. These motions manisfest as vertical oscillations of the stellar disk as measured by SEGUE, RAVE and LAMOST \citep[e.g.,][]{widrow2012galactoseismology,williams2013thewobbly,xu2015ringsand,2017arXiv171003763C,HaiFeng2018,Haifeng2018-2} and may be caused by the passage of a massive satellite such as the Sagittarius dwarf galaxy \citep{gomez2012signatures,gomez2013vertical,widrow2014bending,delavega2015phasewrapping,donghia2016excitation}, although \citet{faure2014radialand}, \citet{Debattista2014}, and \citet{Monari2016} showed that vertical oscillations in the form of breathing modes are also obtained by the passage of the stars through spiral arms. Several independent studies showed that rings, wobbles and in-plane velocity anisotropies  may be the dynamical response of the stellar disk to the gravitational disturbance of a satellite galaxy \citep{Minchev2009,Purcell2011, donghia2016excitation} or their collective effects \citep{Kazantzidis09,Chequers18}. In the case of a perturbation by the Sagittarius dwarf, not only do such disturbances affect the inner part of the galaxy, but they can impact the outer disk in regions where the stellar surface density decreases and the disk is more sensitive to external perturbations. In particular, the recent pre-Gaia DR2 Sgr impact models of \cite{Laporte2017} demonstrated that it could simultaneously account for the morphology of the outer disk and amplitude of density and streaming motion fluctuations in the solar neighbourhood.

Recently, the phase-space spiral was also attributed to the buckling of the Galactic bar \citep{Khoperskov} although \citet{Laporte2018} showed that it affects stars of all ages, which is {\it a priori} more in line with a recent perturbation. It is however possible that perturbations linked to the last pericentric passage of the Sagittarius dwarf and to the buckling of the bar could co-exist in the Milky Way disk.

The extension of the spatial scale of kinematic studies 
to larger regions of the Galactic disk by Gaia DR2 indicated the evidence of streaming motions in the radial, azimuthal and vertical velocity as well as small-amplitude fluctuations in the velocity dispersions in the region between 3 and 8 kpc far from the Solar location. In particular, the data suggest an increase in median vertical velocity, from the inner to the outer disk, that shows a complex dependence on the azimuthal component\citep{Katz2018}.

In this Letter we use an N-body simulation of a Milky Way-sized galaxy to investigate 
the implications of time-varying potentials 
for the phase-space structure and the surface density of the stellar disk induced by the impact of a massive satellite 
galaxy. We show that the assumptions of dynamical equilibrium and a static potential used to infer the disk dynamical density for different azimuthal angles, and particularly at large distances from galaxy center, based on stellar kinematics may not be robust unless 
the time-varying effects are taken into account. 
This paper starts with a brief description of the numerical experiment used in this study and its peculiar phase-space structure (Sec.~2). We then describe the classical Jeans approach that one would be allowed to follow in 
dynamical equilibrium (Sec.~3), and show how it could mislead an observer in deducing the incorrect dynamical surface density of the disk when the potential is actually time-varying.

\section{THE NUMERICAL SIMULATION}
\label{sec:methodology}
\subsection{Set-up}
We analyse a high-resolution N-body simulation of a Milky Way sized galaxy performed with the parallel TreePM
code GADGET-3. A detailed description of the simulation is available in the literature \citep{Laporte2017}.  
In this study the galaxy model consists of a live dark matter halo, a rotationally supported disk of stars,
and a spherical stellar bulge. The dark matter mass distribution of the galaxy is initially modeled with a Hernquist 
profile \citep{hernquist1990ananalytical}. The total mass of the halo is 
$1\times10^{12}{\rm M}_{\odot}$ and is sampled with 40 million particles. The radial scale length of the halo is $52\unit{kpc}$ prior to applying adiabatic contraction according to \citep{blumenthal86}. Each dark particle has a mass of $2.6\times10^{4}{\rm M}_{\odot}$
and a gravitational softening length of $60\,\unit{pc}$. For the stellar disk we adopt an exponential radial 
profile with an isothermal vertical distribution:

\begin{eqnarray}
\rho_{*}(R,z) & = & \Sigma(R)\zeta(z)\nonumber \\
 & = & \frac{M_{\star}}{4\pi z_{0}h_{R}^{2}}\exp\left(-R/h_{R}\right){\rm sech}^{2}\left(\frac{z}{z_0}\right).\label{eq:rho_sech2}
\end{eqnarray}

\noindent
where $M_{\star}$ is the stellar disk mass, $h_{R}$ is the exponential disk scale length and $z_{0}$ is the scale-height of the ${\rm sech}^2$ vertical distribution.
The disk mass is set to $M_{\star}=6.0\times10^{10}{\rm M}_{\odot}$, $h_{R}=3.5 \ \unit{kpc}$, and $z_0=0.53 \  \unit{kpc}$. The stellar disk is discretized with 5 million particles. {\bf The radial and vertical stellar velocity dispersion-squared both follow an exponential profile with the same scale-length as the stellar density $h_\sigma=3.5$~kpc.} The stellar particles have masses of $1.2\times10^{4}\,\unit{M_{\odot}}$ and gravitational softening lengths of $30\,\unit{pc}$.
The stellar bulge is described by the Hernquist model, with a total mass  $1\times10^{10}\,\unit{M_{\odot}}$and
scale radius $0.7 \ \unit{kpc}$.

\subsection{Interaction with a massive satellite}

The satellite adopted in the simulation \citep[Model L2 of][]{Laporte2017} 
has a virial mass of $6 \times 10^{10} M_\odot$ and sinks through dynamical friction for 6~Gyr into the host galaxy halo. It excites a wake in the dark halo, creating torques acting on the disk, before transitioning to acting directly on the disk through tides at late times. As shown in \citet{Laporte2018}, each pericentric passage excites a new phase-space spiral in the $z-v_z$ plane, similar to the one observed in the Solar neighbourhood with Gaia. Each impact induces a prompt response of the disk after the collision, with the stars around the impact region moving inward first and then outwards. A wave of enhanced density is the outcome of this contraction and expansion. Such a wave that propagates through the disk tends to form a two-armed ring-like pattern with over-dense and under-dense regions together with strong bending vertical velocities \citep[see e.g.][]{gomez2013vertical,donghia2016excitation}.

Hereafter, we are studying the problem of performing Jeans modelling on such a galaxy which has generically reacted to a satellite perturbation, in a general fashion, and have chosen a random time of t=6 Gyr for the analysis. The mass model of the reacting galaxy is close to that of the Milky Way in its gross properties, but our Jeans modelling should {\it not} be quantitatively compared to Gaia data, since the model is not meant to be a perfect realization of the present-day Milky Way. However, as already shown in \citet{Laporte2017,Laporte2018}, a number of qualitative and quantitative agreements with the data exist. Here, we divide the disk into six sextants, and display in Fig.1 the breathing and bending velocities in layers of 400~pc height along the vertical component of the disk. Following \citet{Katz2018}, these velocities are computed as the half-sum (mean) and half-difference of the median vertical velocities in symmetric layers with respect to the Galactic mid-plane.

\begin{figure}[th]
\centering
\includegraphics[scale=0.5]{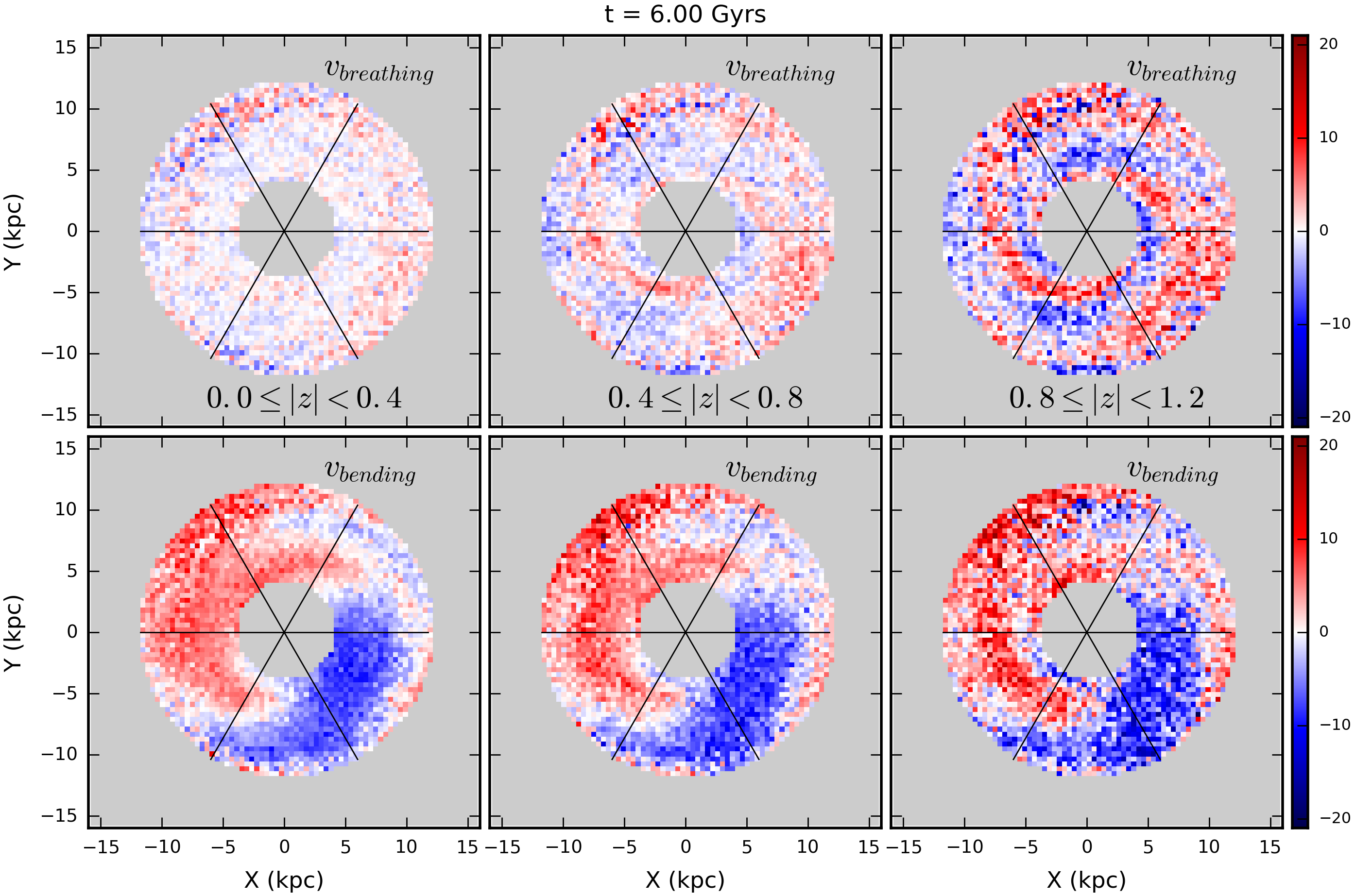}
\caption{\label{fig:vz_modes}}
The breathing (top row) and bending (bottom row) mode velocities
within $8\pm 4$kpc at t$=6.0$ Gyrs. 
\end{figure}

With the above caveat in mind, if we locate the Sun at the sextant with the azimuth angle between $\-60 < \theta < 0$ (which 
corresponds also to an under-dense region in the disk) the bending 
velocities are negative (i.e. they are oriented towards the south Galactic pole) for $|z|$  between [0,400] pc 
and [400, 800] pc but become positive , in qualitative agreement with the Gaia DR2 data. 
Furthermore, the breathing velocity changes from positive to negative values at larger heights from the midplane, 
again in qualitative agreement with the current observations.

At the Solar vicinity, the unprecedented precision and the high-sampling offered by the Gaia DR2 data showed that the stars are distributed in the plane defined by the vertical coordinate and the vertical velocity, $z-v_z$ with a curled spiral-shape whose density increases towards the leading edge of the feature (Antoja et al. 2018). Such a spiral shape is also clearly visible when mapping the average azimuthal velocity as a function of position in the $z-v_z$ surface of section.

This feature is interpreted as the incomplete phase-mixing phenomenon occurring in the stellar disk and a manifestation of the disk being locally out of dynamical equilibrium. Recent studies using analytic models \citep{Antoja2018,BinneySchonrich} or numerical simulations \citep{donghia2016excitation,Laporte2017,Darling19} interpreted this feature as likely caused by the coupling of vertical and in-plane motions induced by the passage of a satellite galaxy through the Galactic disk. In fact in the presence of rings induced by the impact of a satellite,  the intra-ring underdense regions coincide with a local increase of the characteristic scale height \citep[see Fig. 1 in][]{donghia2016excitation}. This is consistent also with the fact that the feature is not traced by the OB stars, which are younger than
100 Myrs and did not complete one orbital period around the Galactic center \citep{Cheng2019}.  In normal conditions for disk stars, the vertical motion of the stars is almost fully decoupled from the in-plane motion with the vertical frequency of stars being usually much greater than the radial epicyclic frequency, with the stars oscillating vertically much faster than their radial motions. However, in under-dense regions, stars feel a weaker gravitational restoring force, and have their vertical frequency reduced. Thus, we expect that stars located in the under-dense regions of the disk will have the horizontal and vertical motions coupled.
Hence the spiral-shape feature in the $z-v_z$ plane should be visible preferentially in the under-dense regions and should vanish more quickly in {\bf the inner parts of the stellar disk} or in locations  with enhanced density like density waves.

Fig. 2 displays the azimuthal velocity in the $z-v_z$ plane
in one of the most underdense regions of the simulated disk at t=6~Gyr, defined by the sextan with the azimuthal angle ranging $-120 < \theta < -60$ (right panel) as compared to the same analysis reported in the enhanced density region in the sextant characterized by $-180 < \theta < -120$ (left panel). Indeed, as expected, the spiral-shape feature is visible in the under-dense region, but has a much less well-defined shape in the over-dense region. {\bf Note that this finding suggests
that the disk self-gravity for time-varying potentials plays a key role in the interpretation of the spiral feature. 
Therefore, approaches based merely on the impulse approximation
and the phase-wrapping of the stellar orbits to describe the feature are limited}.

\begin{figure}[th]
\includegraphics[scale=0.5]{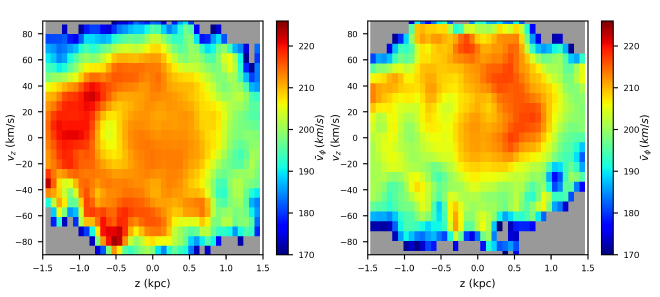}\caption{\label{fig:phase_2d} Two-dimensional surface of section in bins of $\Delta z=0.05$kpc and $\Delta v_z=1$km/s colored as a function of the average $v_\phi$ for stars within 500pc of $R_\odot=8$kpc and $-1.5\leq z \leq 1.5$kpc at t=6Gyrs in the $-180 \leq \theta < -120$ azimuthal bin (left) and the $-120 \leq \theta < -60$ bin (right) smoothed using a Gaussian with zero mean and $\sigma=1$.}
\end{figure}

\section{The Jeans Analysis}
\label{subsec:analysis}

We now proceed with analyzing the simulation at two different times, namely (i) $t=0.75$~Gyr, after the disk has settled from the initial conditions, but before the massive satellite has had a significant impact on it, and (ii) $t=6$~Gyr, an arbitrary time-step 100 Myr before the last pericentric passage of the massive satellite.

In both cases, we make a vertical Jeans analysis at $R=8$~kpc by constructing an annulus 200 pc in width at this radius. This annulus is then broken into equally-sized azimuthal patches that approximately correspond to the view used by \citet{Katz2018} to analyze the vertical velocity field of the \textit{Gaia} DR2 data (see Fig.~1).

As is traditionally done, we then combine the vertical Jeans equation with the Poisson equation to obtain an estimate of the dynamical surface density as a function of height at $R=8$~kpc. This Jeans estimate can then directly be compared with the actual surface density of stars and dark matter in the simulation.

The equation we use for the Jeans estimate of the dynamical surface density at $R=8$~kpc is \citep[see, e.g.,][]{HagenHelmi2018}

\begin{eqnarray}
2\pi G\Sigma_{dyn}\left(R,z\right) & \simeq & \frac{\sigma_z^{2}}{h_{z}}-\frac{\partial\sigma_z^{2}}{\partial z}\label{eq:SigmaTot}\\
 &  & -{\rm cov}\left(v_{R},v_{z}\right)\left[\frac{1}{R}-\frac{2}{h_{R}}\right]\nonumber \\
 &  & +2 v_{c} \frac{|z|}{R} \frac{\partial v_{c}}{\partial R}\nonumber 
\end{eqnarray}

\noindent where the three first terms correspond to the estimate of the vertical force from the vertical Jeans equation and the last term accounts for a non-flat rotation curve at the radius of interest, assuming that it does not vary significantly within the height of interest. Here, $h_z = - ({\rm d ln}(\rho)/{\rm d}z)^{-1}$ {\bf is just a notation corresponding to the local ``exponential scale-height" at the height of interest}, not to be confused with the (constant) scale-height $z_0$ of Eq.(1). {\bf We note that $h_z$ is calculated independently in each azimuthal bin, as well.} {\bf The vertical density profile is not exponential. Note that this equation is based on a few important assumptions, notably that the radial and vertical velocity dispersion-squared vary with radius with the same exponential scale length as the density, and that the velocity ellipsoid tilt angle is locally constant with radius. The assumption that the dispersion-squared follows the same radial exponential profile is justified by the set-up of the initial conditions. As we shall see below, this set of assumptions indeed yields reasonable results when the disk is unperturbed (Fig.~3). However, an obvious other assumption is that the disc is axisymmetric indeed. In this case, while the results at $t=0.75$~Gyr should logically not be affected, it is not completely clear how much this assumption will affect the results at $t=6$~Gyr (see Sects.~4 and 5).}

\begin{figure}
\includegraphics{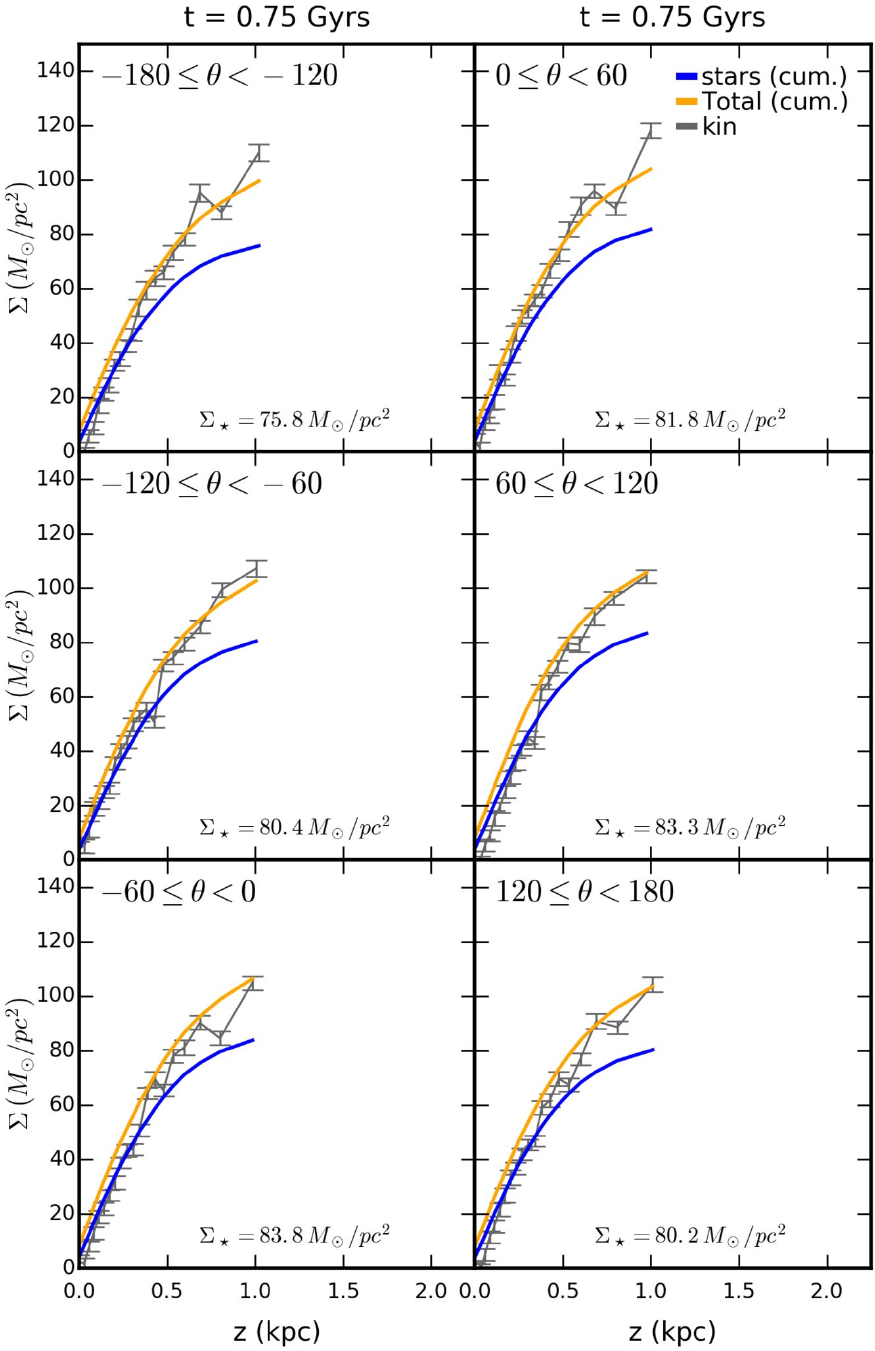}
\caption{\label{fig:smd_z_0_75}Cumulative stellar surface mass density(blue),
cumulative total density (stars and dark matter; orange),
and the density derived from Eqn. \ref{eq:SigmaTot} (grey)
as a function of height above the midplane where the lower half-plane
has been folded onto the upper half-plane in each azimuthal bin at t=0.75 Gyrs.}
\end{figure}

Within each azimuthal patch, we recenter the stars by subtracting
the position of the peak vertical density. The stars are then split into
vertical bins of equal counts to provide an approximately equal signal-to-noise in each bin when measuring the velocity dispersion. The vertical density is fit by a ${\rm sech}^2$ distribution, which allows us to compute $h_z(z)$. At $t=0.75$~Gyr, the simulated disk is nicely plane-symmetric, which allows us to fit a single ${\rm sech}^2$ to the whole distribution and test our framework by computing the total density inferred from Eq.(\ref{eq:SigmaTot}), $\Sigma_{dyn}$, and the actual total density, and compare it to $\Sigma_{{\rm tot}}=\Sigma_{\star}+\Sigma_{DM}$. The results are shown in Fig.~\ref{fig:smd_z_0_75}, and demonstrate that the method works reasonably well when the simulated disk is close to equilibrium.

At $t=6$~Gyr, even after correcting for the position of the peak vertical density, the important North-South asymmetries in the vertical distribution of stars force us (as an observer confronted to such asymmetries would be) to consider a separate Jeans analysis in the Northern and Southern Galactic hemispheres.

\section{Results}

\begin{figure}
\includegraphics{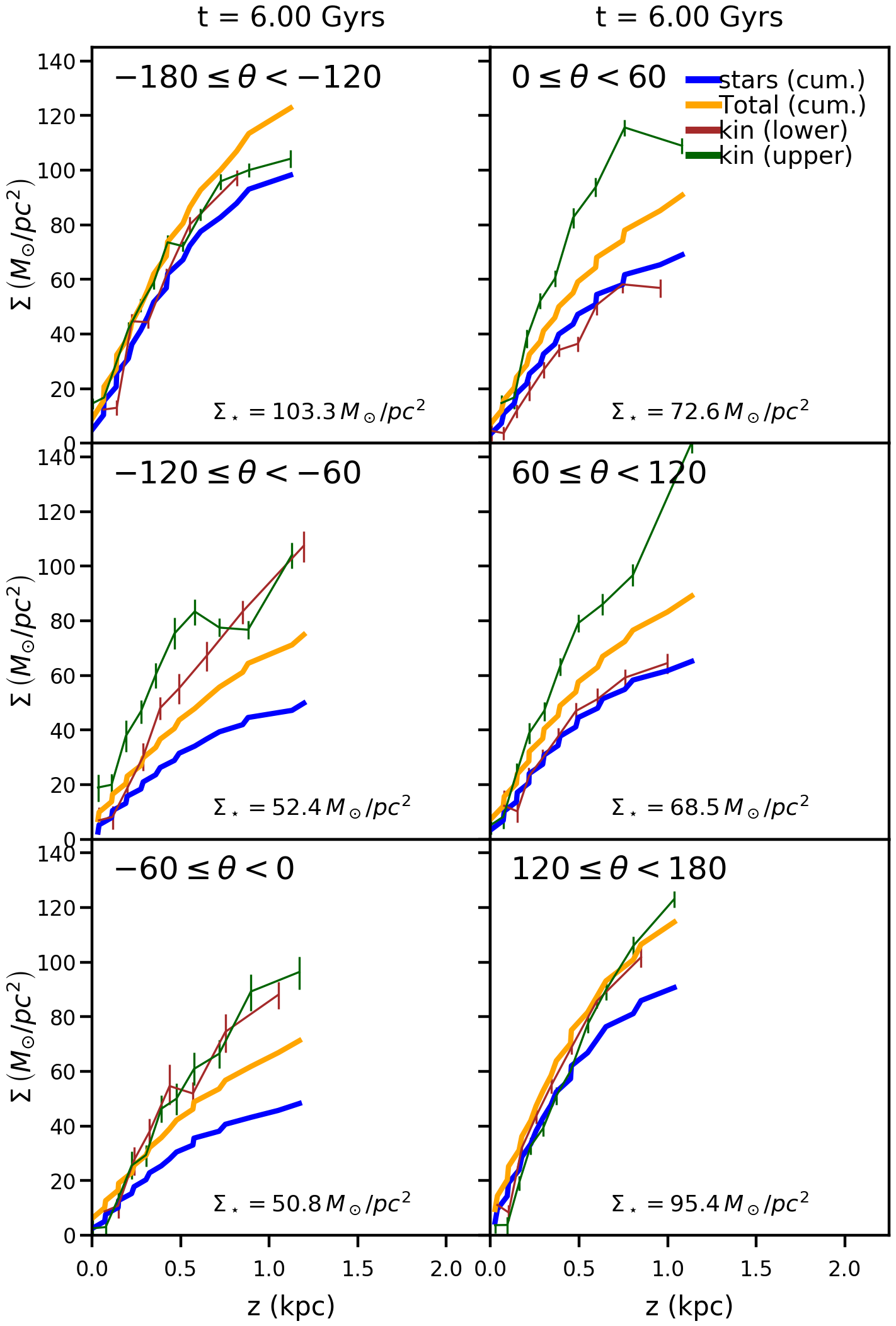}
\caption{\label{fig:smd_z_6_00}Cumulative stellar surface mass density(blue),
cumulative total density (stars and dark matter; orange),
and the density derived from Eq. \ref{eq:SigmaTot} for z$<0$kpc (brown-red) and z$>0$kpc (green)
as a function of height above the midplane in each azimuthal bin at t=6 Gyrs.}
\end{figure}

Fig. \ref{fig:smd_z_6_00} shows the cumulative stellar surface mass density (SMD) (blue), the cumulative total matter SMD (stars + DM; orange), the SMD derived from Eq. 2 using the stellar kinematics (green at $z>0$ and red at $z<0$) as described in Section 2.1 as a function of height above the plane. 

We note that, in the most over-dense regions of the stellar disk (the upper left and bottom right panels), the matter density dynamically inferred by the Jeans analysis tends to be consistent in both the Northern and Southern hemispheres, and is consistent with the total matter density present in the simulations (albeit with a slight underestimation at large heights in the sextant $-180<\theta<120$). On the other hand, in the most under-dense regions of the stellar disk (the left middle and bottom panels) the matter surface density inferred from the Jeans analysis is also quite consistent in both hemispheres, but systematically overestimates the total surface density. This systematic overestimation is in line with the previous analysis 
of a simpler model of a disk affected by breathing modes (see Figure 6 in \citealt{banik2017galactoseismology}). 
Finally, in the regions of average density, the Northern and Southern estimates tend to disagree with each other, but their average is close to the true value. {\bf For patches where the North and South Jeans analyses give consistent answers, one might wonder whether the assumptions of Eq.~2 might be the culprits. The stellar density scale-length is not modified at 6~Gyr compared to the initial conditions, once averaging over the wiggles due to azimuthal over- and under-densities. The velocity dispersion-squared scale length is largely unmodified too, although it varies slightly from one azimuth to the other, between one and twice the density scale-length. This, together with the assumption on the tilt angle \citep[see, e.g.,][for recent discussions]{Hagenveleps,Evans}, could modify the results slightly. Moreover, due to the clear non-axisymmetry at 6~Gyr, an axial term should also be included \citep{silverwood2016anonparametric}. However, we expect all these terms to be largely subdominant with respect to the first term of the r.h.s. of Eq.~(2), which largely dominates the result we obtained, as the last two terms only account for modifications at the level of $\sim 1 M_\odot/pc^2$. Indeed, we can interpret the overestimate of the dynamical surface density in underdense regions mostly in terms of $h_z=({\rm d ln}(\rho)/{\rm d}z)^{-1}$, which is too small compared to its equilibrium value in those underdense regions. Moreover, although unlikely, even if a thorough non-axisymmetric analysis would be able to recover the correct surface density, it could never explain the large North-South differences found in 2 out of 6 sextans, which, if found in a data analysis, would be a smoking gun of disequilibrium. Interestingly, the actual stellar densities are actually rather symmetric in the North and South (with differences only of the order of $\sim$10~$M_\odot/pc^{2}$ in those patches where the derived surface densities are strongly asymmetric.} 

\section{Conclusions}

In this work, we extended the classical Jeans analysis to a simulated stellar disk impacted by a massive satellite. This collision induces a vertical wobble in the disk and in-plane rings, features that are not accounted for in the estimate of the local matter density inferred from the observed kinematics of the stars. {\bf Our analysis is a simple zeroth-order one, as is often performed in the solar neighbourhood, without using non-parametric approaches or taking into account the possible effects of non-axisymmetry \citep{Widmark2019, Widmark2, silverwood2016anonparametric}. }

Our main finding is that, while this simple Jeans modelling gives reasonable estimates of the total surface density in the most over-dense regions of the disk, it tends to systematically overestimate the surface density, by a factor as large as 1.5, in the most under-dense regions, {\bf where the scale-height is too small compared to its equilibrium value}. This reinforces our interpretation that the vertical motion of the stars is coupled to their in-plane motion in these under-dense regions, thereby rendering the equilibrium analysis obsolete. We also point out that, in the regions of average density, the Jeans analysis tends to give very discrepant results in the Northern and Southern Galactic hemispheres, whose average value is nevertheless close to the true one. {\bf Such North-South asymmetries are a smoking gun of disequilibrium, as a thorough non-axisymmetric analysis, by construction, could never account for such asymmetries.}

Our findings are highly relevant for future attempts at determining the dynamical surface density of the outer Milky Way disk as a function of radius, by combining Gaia data with future spectroscopic surveys. The important biases of simple Jeans modelling, especially in under-dense regions call for the development of non-equilibrium methods to estimate the dynamical matter density in the outer disk. For instance, the method recently proposed by \citet{BinneySchonrich} could make it possible to take advantage of the presence of a phase-space spiral to constrain the potential once the timing of the main perturber (for instance the last pericentric passage of the Sgr dwarf) is known. However, we note that such methods rely on the impulse approximation, whereas yet again indicated by our results the self-gravity of the disk is relevant, meaning that the method should be improved to take it into account.

\section*{Acknowledgements}

{\bf We thank the anonymous referee for a thoughtful and constructive report.} We are grateful to Justin Read and George Lake for their valuable
advice. E.D.O acknowledges support from the Vilas Associate Research Fellowship and thanks the Institute for Theory and Computation (ITC) at Harvard  and the Center for Computational Astrophysics at the Flatiron Institute for the hospitality during the completion of this work. CL is supported by a CITA National Fellow award. This work is supported by ATP NASA Grant No NNX144AP53G. This work used the Extreme Science and Engineering Discovery Environment (XSEDE), which is supported by National Science Foundation grant number OCI-1053575. E.D.O, B.F and C.L acknowledge hospitality at the KITP, supported by the National Science Foundation under Grant No. NSF PHY-1748958, in the final stages of this work.

\bibliographystyle{apj}
\bibliography{OortLimit}

\end{document}